\documentstyle[sprocl]{article}

\def\Journal#1#2#3#4{{#1} {\bf #2}, #3 (#4)}

\def\NPB{{\em Nucl. Phys.} B}
\def\PLB{{\em Phys. Lett.}  B}
\def\PRL{\em Phys. Rev. Lett.}
\def\PRD{{\em Phys. Rev.} D}
\def\ZPC{{\em Z. Phys.} C}

\def\be{\begin{equation}}
\def\ee{\end{equation}}
\def\bea{\begin{eqnarray}}
\def\eea{\end{eqnarray}}

\begin{document}

\title{REVIEW OF THEORY TALKS AT XXIX INTERNATIONAL SYMPOSIUM ON
MULTIPARTICLE DYNAMICS}

\author{A. KAIDALOV}

\address{Institute of Theoretical and Experimental Physics,
B. Cheremushkinskaya Street 25,
Moscow,\\ 117259, Russia\\E-mail: kaidalov@vxitep.itep.ru}


\maketitle
\abstracts{ A short summary of main results of theoretical talks
presented at XXIX International Symposium on Multiparticle Dynamics is
given.}

\section{Introduction}

The main trend in theoretical applications of QCD to processes of
multiparticle production during recent years, which was well presented at
 this Symposium, can be summarised as:\\
"From small to large distances or an interplay of perturbative and
nonperturbative aspects of QCD".

The basis of QCD is rather well established now by a comparison of its
predictions with experiments sensitive to small distances (large
 virtualities or momentum transfer), which can be described using the
QCD perturbation theory.

On the other hand a large distance dynamics is still an open problem
and we can not claim that we understand QCD without solving it. So
main efforts have been concentrated on investigation of different
dynamical aspects of QCD for a broad variaty of phenomena of
 multiparticle production. One of the crucial problems is to understand
what are the relevant degrees of freedom in different processes?
Are they point like quarks and gluons or "reggeized" quarks and gluons
or rather white objects,- Pomeron and reggeons?

 Plan of my talk is related to the problems mentioned above. In
the first Section I shall concentrate mainly on small distance processes and
physics of jets. In Section 2 the problem of Pomeron and its
 manifestations in diffractive processes and multiparticle production
will be reviewed. There was a substantial progress in this field during
last 2 years and in more than 50\% of theoretical talks at this Symposium
different aspects of this problem have been discussed. In particular
in the third Section I shall consider shadowing effects in small-x physics
and their relation to heavy ion interactions, though I shall not discuss
the field of heavy ion interactions in details as it was perfectly reviewed
by J.Stachel~\cite{Stachel}. Other theoretical ideas and their applications
to multiparticle production processes will be discussed in Section 4.

I would like to apologize to many speakers of this Conference for being
unable to cover in this talk their interesting contributions and possible
misinterpretations of some results included in my summary.

\section{Perturbative QCD, jets and power corrections}

Large distance dynamics is present in all physical processes, even in
such typical "small distance" reactions as $e^+e^-$-annihilation at large
energies or deep inelastic scattering. The factorization property in QCD
allows one to separate contributions from small and large distances. For
example cross section for production of jets in hadronic collisions can
be expressed as a sum of convolutions of partonic distributions in colliding
hadrons with corresponding hard cross sections. These cross sections can be
calculated in QCD perturbation theory. A dependence of partonic
 distributions on a scale $\mu$ of a process can be determined using the
renormalisation group equations and can be described perturbatively
at large $\mu$, while initial conditions (values of partonic distributions
at fixed scale $\mu_0$ ) are determined by both small and large distance dynamics
and in general can not be predicted by perturbative QCD. Due to confinement
of quarks and gluons in QCD they are observed as jets of hadrons and a
transition from partons to hadrons is a necessary step in theoretical
calculations.

Impressive agreement of perturbative QCD with experimental data has been
 demonstrated at this Symposium. HERA data on the proton structure function
 $F_2$ can be well described by the QCD evolution equations in a broad region
 of $Q^2$~ \cite{Zhokin} and provide an information on distributions of quarks
 and gluons at very small x.

 Cross sections of jets production are in an agreement with PQCD calculations
 both at Tevatron \cite{Seidel,Babuk}, HERA \cite{Hadig} and $e^+e^-$ annihilation
 \cite{Seibel,Abe,Sarkar}.
 Infrared safe characteristics of jets are
well described by perturbative QCD if power corrections are taken into account
(see below).

 A substantial progress in separation of quark and gluon jets has been achieved
in recent years and a trend to the asympototic prediction of QCD for multiplicities
of these jets
\be
\frac{\bar n_g}{\bar n_q}=\frac{C_A}{C_F}=\frac{9}{4}
\ee
is confirmed \cite{Seibel}.

Different aspects of nonperturbative physics of jets hadronization have
been discussed in several talks at this Symposium. It was shown by
 S.Chun \cite{Chun} that the model based on area law gives a good description of
 relative yields of different hadrons. An importance of spin-spin interactions,
for particle production has been emphasized by P.Chliapnikov \cite{Chliap}.
A role of nontrivial color connections between different partons, which lead to
$1/N_c^2$ corrections to the leading planar configurations, has been studied
by Q.-B Xie \cite{Xie}. A model based on stachastic branching and local parton
-hadron duality hypothesis(LPHD) was developed by A.H.Chan \cite{Chan}.

The LPHD hypothesis has been formulated many years ago by Azimov et al. \cite{Azimov}
and found to be very useful for description of some global properties of
multiparticle production in hard processes. An interesting new application of this
LPHD was reported at this Conference by W.Ochs \cite{Ochs}. He calculated a probability
of events in $e^+e^-$-annihilation with a rapidity gap. For partons such events
with absence of partons in a large region of $\Delta y$ are suppressed by Sudakov
factor. If LPHD is correct suppression of the same type should exist also for
hadrons.

It is clear that LPHD can not be true in all situations and it is very important
to understand why it works and where it fails?  First experimental indications
to violations of LPHD has been discussed in several talks at this Symposium
 \cite{Seibel,Milstead,Metzger}.

Hadronisation effects in properties of jets are closely related to power
corrections to these quantities, which were found to be essential for
accurate description of jets observables. This problem clearly reflects an
interplay between PQCD and large distance physics. There is a hope that
these corrections can be described in terms of a quantity $\alpha_0(\mu)$,
which is an average value of the strong coupling $\alpha_s(k^2)$ in the
 region of small virtualities $k^2<\mu^2$~ \cite{Dokshit}. Such approach is
useful only if  $\alpha_0(\mu)$ is universal for different observables.
These corrections were found to be approximately universal for jets observed
at LEP \cite{Sarkar} and at HERA \cite{Milstead}, though HERA experiments
indicate to some differences between values of  $\alpha_0(\mu)$ from
different observables. In my opinion an agreement is better than one can expect
for such a simple model of power corrections.

Perturbative approach to production of events with large rapidity gaps between
jets in hadronic collisions has been discussed by G.Sterman \cite{Sterman}.
 He proposed to characterize a hadronic activity in rapidity region
between jets by energy
flow $Q_c$ and have shown that it is possible to describe by perturbative QCD
the region where $Q_c$ is much smaller than $p_T$ of jets but larger than
characteristic hadronic scale.

Cross sections of semihard interactions in hadronic collisions fastly increase
with energy and multiparton interactions become important at superhigh
energies. This problem has been addressed in several
 talks \cite{Trel,Calucci,Pancheri,Ugoccioni,Walker}.

Properties of multiparton distributions have been discussed by D.Treleani and
G.Calucci \cite{Trel,Calucci}. These distributions are characterized by ratios
of momenta $x_i$ and impact parameters $b_i$ of all partons. It was pointed out
that simplest uncorrelated distribution fails to reproduce experimental data
on cross section of double parton interaction and possible correlations
between valence quarks and gluons can lead to an agreement with experiment.

Applications of models with multiple partonic interactions to total interactions
 cross sections in h-h,$\gamma-h$ and $\gamma-\gamma$ collisions have been
 considered by G.Pancheri \cite{Pancheri} and to multiparticle production by R.Ugoccioni
 \cite{Ugoccioni}. W.D.Walker \cite{Walker} has demonstrated that multiple
 interactions are needed for understanding of Tevatron data on multiplicity
 distributions.

Impressive collection of new results on interactions of real and virtual photons
has been presented by LEP experiments \cite{Feld,DeRoeck}. A fast increase
with energy of these cross sections is observed at highest energies. It is
difficult to reconcile the increase observed by L3 group for $\gamma-\gamma$
interactions with theoretical models based on an eikonalized version of mini-jet
model \cite{Pancheri}. First results on interaction of highly virtual photons
provide a good testing ground for recent predictions based on NLO BFKL
Pomeron calculations \cite{Kim}.

Spin effects in processes of interaction of a virtual photon with proton have
been discussed by N.Nikolaev \cite{Nikolaev}. He have shown that simplest diagrams of two
gluon exchange for vector meson production by highly virtual photons leads to
the spin structure of $\gamma^* V$ transition, which is in an agreement with
HERA data (see also \cite{Ivanov}). In particulat the model leads to a definite
pattern of violation of s-channel helicity conservation. It was also shown in this talk
that due to double-Pomeron exchange structure function $g_2$ has a singular
behaviour as $x\to 0$. As a result the Burkhardt-Cottingham sum rule and
Wandzura-Wilczek relations are violated.

\section{Pomeron}

The notion of the Pomeron has been introduced in particle physics in the
framework of Regge theory long ago \cite{Pomeron}. There is a revival of
 interest to the Pomeron problem due to small-x physics studied at HERA.
 The Pomeron plays an important role in theoretical descriptions of
high-energy interactions, however there are no unique definition of this object. So
I shall first discuss existing definition of the Pomeron. They can be
devided into two categories:\\
a) Pomeron is the Regge pole with the largest intercept $\alpha_P(0)$ and
vacuum quantum numbers. It gives a contribution to high-energy amplitudes
 of elastic scattering and other diffractive processes. In this approach
 multipomeron exchanges, which lead to moving cuts in the complex angular
 momentum plane $j$, exist also. They are especially important in the case of
 "supercritical" Pomeron (when $\alpha_P(0)>1$) to restore unitarity of the
 theory and to satisfy Froissart limit.\\
b) Pomeron is the singularity at $j=1$ (not a Regge pole in general), which
satisfies constraints of unitaruty, analyticity  and describes asymptotically
diffractive processes.

 I prefer the first definition due to the following reasons:\\
i) It relates high-energy scattering to hadronic spectrum.\\
ii)It is natural in $1/N$-expansion in QCD.\\
iii) Multiparticle content of the Pomeron is known (short range correlations).\\
iv) Gribov reggeon diagrams technique allows one to estimate amplitudes for
 multipomeron exchanges  and AGK (Abramovsky, Gribov, Kancheli) cutting rules
 relate them to multiparticle production processes. Such approach leads to
 a successful phenomenology \cite{review}.

It is very important to understand dynamics of reggeons and of the Pomeron
in QCD. A useful framework to classify all diagrams in QCD is
 1/N-expansion \cite{1/N}, where N is either number of colors $N_c$
or light flavors
$N_f$. In this approach the reggeons $\rho,\omega,A_2,$ are connected to
planar diagrams, while the Pomeron is related to cylinder type diagrams.
Diagrams with exchange by $n$ Pomerons in the t-channel are connected to
 multicylinder configurations, which are $\sim (1/N^2)^n$ and are small in
the large $N$ limit. In realistic calculations these contributions should be
taken into account.
Such classification leads to many predictions for high-energy hadronic
interactions, which are in a good agreement with experiment \cite{review}.

 Calculation of reggeon and Pomeron trajectories in QCD with an account of
nonperturbative effects is a difficult problem (for some recent results
see below). Perturbative calculations of the Pomeron in QCD have been
carried out by L.Lipatov and collaborators \cite{BFKL} (BFKL Pomeron) many
years ago. Pomeron is related to a sum of ladder type diagrams with
 exchange by reggeized gluons. Reggeization of gluons (as well as quarks)
is an important property of QCD (at least in perturbation theory).
 In the leading approximation an expression for the intercept of the
Pomeron is well known \cite{BFKL}
\be
\Delta\equiv \alpha_P(0)-1=\frac{4N_cln2}{\pi}\alpha_s
\ee
In this approximation it is not clear which value of $\alpha_s$ to use and
for  $\alpha_s=0.2$  an intercept of the Pomeron is substantially above
unity $\Delta\approx 0.5$. It leads to a fast increase of total cross
sections $\sim (s/s_0)^{\Delta}$ with energy. The arguments were given
 that next to leading corrections to the Pomeron intercept should be
 large \cite{Anderson,Kancheli}. This is connected to the observation of
relatively small average rapidity intervals in the gluon ladder for realistic
values of $\alpha_s$, while LO expressions are valid for large rapidity
intervals. NLO corrections have been calculated last year \cite{Fadin,Camici}
and strongly modify LO results for $\Delta$
\be
\Delta=2.77\alpha_s(1-6.5\alpha_s)
\ee
For $\alpha_s>0.15~~\Delta$ becomes negative. It is clear that an origin of
large NLO corrections should be clearly established and resummation of
these effects is necessary.

Some results in this direction were presented at this Conference. It was
pointed out by V.Kim \cite{Kim,Brodsky} that results for $\Delta$ depend on
a choice of renormalization scheme and renormalization scale
(original result was obtained in $\bar{MS}$-scheme). The choice of more
physical (BLM) scheme leads to more stable results for $\Delta$, which
 practically does not depend on $Q^2$ and $\Delta\approx 0.17$.

Another approach has been developed by M.Ciafaloni et al. \cite{Ciafaloni}
and was presented by D.Colferai \cite{Colferai}.  They perform a partial
 resummation of subleading corrections using renormalization group analysis.
In the case of two scales processes (like DIS) an intercept of the "hard
Pomeron" $\alpha_P(Q^2)$ is introduced and investigated. This quantity can
determine behaviour of structure function $F_2$ for large $Q^2$ and not too
small x. It is pointed out that an intercept of the leading Regge pole $\alpha_P$
(which is of course should not depend on $Q^2$) depends on dynamics in
nonperturbative region. The Pomeron problem is a clear manifestation of an
interplay of soft and hard mechanisms in QCD.

An importance of nonperturbative effects and especially of chiral symmetry
breaking effects for dynamics of the Pomeron was emphasized by A.White
\cite{White}. He uses a powerful tool of reggeon unitarity for investigation
of interactions of reggeized gluons and quarks in QCD. An important role of
the special U(1) anomaly was demonstrated.

New factorization formula for high-energy scattering amplitudes was obtained by
Ya.Balitsky \cite{Balit}. It allows one to formulate an effective action, which
can be used for calculation of higher order perturbative corrections to BFKL
Pomeron and unitarization effects. This approach is effective for small coupling
$\alpha_s$ and large fields. Related approach has been developed by L.McLerran
with collaborators \cite{McLer} and was reported at this Conference in talks
of J.Jalilian-Marian \cite{Jalil} and R.Venugopalan \cite{Venug}.

It is very important to understand a role of multigluon exchanges for asymptotic
behaviour of scattering amplitudes. This problem has been studied in the
eikonal approximation by H.Fried \cite{Fried}.

Interesting attempt to calculate spectrum of glueballs using metods
developed in the superstring theory has been presented by R.Brower
\cite{Brower}. The leading Regge trajectories of the glueball spectra can
be related to the Pomeron Regge pole. Recently Yu.Simonov and myself have
calculated spectrum of glueballs using method of vacuum correletors
\cite{Simonov}. Predicted
masses of the lowest glueballs are in a perfect agreement with lattice
calculations. We emphasize an importance of mixing between gluons and
quarks in the low t-region. The mixing effects allow to obtain
phenomenologically acceptable intercept of the Pomeron trajectory and
lead to an interesting pattern of vacuum trajectories in the positive
t-region. I think that the Pomeron in QCD has a very rich and interesting
dynamics.

\section{Shadowing effects in small-x region and "hard" diffraction}

Experiments at HERA clearly demonstrated a fast increase of densities
of quarks and gluons as x decreases. For very large densities partons
will interact and shadow each other. This will lead to a suppresion of the
fast rise of parton densities and finally to saturation of parton densities
 as $x\to 0$.

Same effects can be viewed in the target rest frame as a result of coherent
 multiple interactions of the initial quark-gluon fluctutation of a virtual
photon with the target (note that a fluctuation of a virtual photon with
 small x has a very long lifetime $\tau\sim 1/mx$). In the framework of the
reggeon theory these rescatterings correspond to multipomeron exchanges
in $\gamma^*p (\gamma^*A)$ elastic scattering amplitudes. Investigations of
a role of multipomeron exchanges for dense parton systems have been discussed
by E.Gotsman \cite{Gotsman} and B.Gay Ducati \cite{Ducati}. An equation, which
includes all multipomeron exchanges in the double logarithmic approximation
has been obtained \cite{Ducati} from the dipole picture. It coincides with AGL
equation obtained ealier using Glauber-Mueller approach and can be considered as a
 candidate for unitarized evolution equation at small x. Effects of the
rescatterings (or screening corrections) on the structure function $F_2(x,Q^2)$
have been considered in details by E.Gotsman and the problem of saturation
for parton densities was discussed.

Rescatterings in reggeon theory are closely related to diffractive production
(large rapidity gaps). Experimental results of HERA on these processes have
been discussed at this Symposium by A.Zhokin \cite{Zhokin}
 and K.Piotrzkowski \cite{Piotr}
There was a considerable interest in the processes of "hard" diffraction in
recent years. If the Pomeron is a factorizable object than one can introduce
the Pomeron structure function, which characterize a distribution of quarks
 in the Pomeron $F_P(\beta,Q^2)$. Experiments at HERA found that effective intercept
of the exchanged object for large rapidity gap events $\Delta_{eff}=0.15\div 0.2$
 at large $Q^2$. This value is larger than corresponding values in soft
diffraction. Structure function of the Pomeron was also determined by H1 and
ZEUS \cite{Zhokin}.

The model for distribution of quarks and gluons in the Pomeron has been
considered by F.Hautmann \cite{Haut}. It is based on perturbative QCD
approach to this problem with the assumption of dominant role of small transverse
sizes for initial distribution of quarks and gluons in the Pomeron. Dependence
on $Q^2$ was calculated using standard QCD evolution.

Note that for inelastic diffraction multipomeron exchanges are also present
(and even more important than for elastic amplitudes) and in general
amplitudes of these processes are not factorizable. A simultaneous
 sefconsistent description of both $F_2(x,Q^2)$
in a broad region of $Q^2$ and diffractive production $F_2^{D3}(x,Q^2,\beta,x_P)$
is a difficult problem. First results in this direction were presented
at this Symposium \cite{Kaid}.

The problem of "saturation" at large $ Q^2$ and x much smaller than those
available at HERA is still not solved completely.
 The region of $ln(1/x)$ and $Q^2$, where the saturation
 happens should be well defined and the question whether $\sigma_{\gamma^*p}$
is large( $\sim Const$) or it is still small ($\sim 1/Q^2$) should be solved.

Results on hard diffraction at Tevatron have also been presented at this
Conference \cite{Convery,Barreto}. Diffractive production of jets, W-bosons,
b-quarks and $J/\psi$-mesons is observed at $\sim 1\% $ level. These
signals are $5\div 10$ smaller than expected from Regge factorization. This
damping factor is expected due to large shadowing effects for inelastic
diffraction in hadronic collisions. A similar suppression takes place for
total cross section of diffraction dissociation. Theoretical estimates
show that these shadowing effects due to multipomeron exchanges influence
mostly total rate and s-dependence of diffractive processes, but have a
little effect on mass or $\beta$-dependence. From this point of view an
observation of very fast increase in diffractive production of jets at
very small $\beta$ observed by CDF group \cite{Convery} looks very interesting.
It contradicts to the parametrisation of distribution for gluons in the
Pomeron proposed by H1. Same comparison should be done for other
parametrisations of gluons proposed in literature. Note that direct information
on small beta behaviour of partonic distributions in the Pomeron (especially
for gluons) is practically absent at HERA.

An important testing ground for Regge factorization and its vilolation is
provided by the process of central production of jets, heavy quarks, e.t.c
in hadronic interactions with two large rapidity gaps (double Pomeron
exchange). Experimental information on this process is still very limited
and more data are clearly needed.

Experimental information on diffraction production, including hard diffraction,
can be understood using the prescription of "flux renormalisation", introduced
by K.Goulianos \cite{Goul}. At present it does not have clear theoretical
basis and it is necessary to understand why it works in many situations.

Another explanation of "Dino's paradox" has been proposed by Chung-I Tan
\cite{Tan}. He emphasizes a role of "flavouring" of the Pomeron, which accounts
for preasymptotic effects due to delayed thresholds of heavy states production.
In this approach it is possible to describe a slow rise of $\sigma^{SD}$ at
very high energies. It would be interesting to see how this approach reproduces
main observational facts for hard diffractive processes.

 Shadowing for dense parton systems in the small x region are especially
important for nuclei, where density of partons for given impact parameter is
larger by a factor $A^{1/3}$. Nuclei are also convenient for a study of these
effects as they can be easily extracted by a study of A-dependence of nuclear
structure functions. This problem has been discussed in several
 talks \cite{Venug,Jalil,Kaid,Sarc}. Though the models considered in these talks
are rather different their predictions look similar. In particular for
heavy nuclei $ (A\approx 200), Q^2\sim 5 GeV^2$ and $x\sim 10^{-4}$ there is a
suppression factor $0.5\div 0.6$  due to shadowing.

This result is important also for heavy ion collisions at RHIC and LHC as it
reduces density of produced minijets (and hadrons).

Same effects were considered in the model of string fusion in the talk of
M.Braun \cite{Braun}. He discussed a possible phase transition due to
 percolation of strings and its influence on fluctuations in heavy ion collisions.

\section{Models for multiparticle production and phenomenological applications}

In this section I shall consider some new developements in models of
multiparticle production and applications of existing theoretical ideas
to different aspects of high-energy interactions.

 The model of color mutations with self-similar dynamics for particle production
in soft processes has been discussed
by R.Hwa \cite{Hwa}. A general organisation of diagrams is similar to the
one used in $1/N$-expansion approach,-the Pomeron corresponds to the
cylinder contribution and multipomeron exchanges in the eikonal approximation
are also taken into account in this model. Dynamics of multiparticle
production for a single cylinder differs from string models. It
is especially important for local (in rapidity) properties of particle
production. The model reproduces experimental data on intermittency, which
pose a problem for existing string models.

Applications of existing models based on $1/N$-expansion, reggeon theory
and string dynamics to cosmic ray physics have been presented by
 R.Engel \cite{Engel}. Comparison of predictions of these models with existing
cosmic ray data indicate to possible problems of existing models at
superhigh energies.

The Pomeron in perturbative QCD is related to exchange by even number of gluons
 in the t-channel. Exchanges by odd number of gluons lead to a singularity
in $j$-plane with negative signature and C-parity, which is usually called
"odderon". Recent perturbative calculations established that in
 LO approximation intercept of the odderon is below unity, but very close
to it. Experimental observation of manifestations of odderon would be an
important check of perturbative QCD predictions (note that lattice
calculations indicate that nonperturbative glueball trajectories of this type have
a very low or even negative intercept). It was shown in the talk of C.Merino
\cite{Merino} that an asymmetry in distribution of charm jets produced in diffractive
photoproduction is sensitive to odderon contribution.

Interesting applications of small-x QCD physics to superhigh energy $\nu N (\nu A)$
interactions and attenuation of $\nu$  transversing the Earth have been
discussed by A.Stasto \cite{Stasto}. Such calculations are important for
$\nu$-astronomy as well as for investigation of atmospheric neutrinos.

 I think that this Symposium demonstrated that
our field of QCD studies in processes of multiparticle production is very
rich and active. Most topical problems now are related to connection
between soft and hard dynamics in QCD. There are many interesting
relations between different fields like small-x DIS and heavy ion collisions.
New experiments at RHIC and later at LHC will give a new impact to
this field of research.

I would like to thank organizers of this Symposium and especially
Chung-I Tan for invitation to participate and to give theoretical review
talk at the Symposium.


\section*{References}
\begin{thebibliography}{99}
\bibitem{Stachel} J.Stachel, Talk at this Conference.
\bibitem{Zhokin}   A.Zhokin, Talk at this Conference.
\bibitem{Seidel} S.Seidel, Talk at this Conference.
\bibitem{Babuk} L.Babukhadia, Talk at this Conference.
\bibitem{Hadig} T.Hadig, Talk at this Conference.
\bibitem{Seibel} M.Seibel, Talk at this Conference.
\bibitem{Abe} T.Abe, Talk at this Conference.
\bibitem{Sarkar} S.Sarkar, Talk at this Conference.
\bibitem{Chun} S.Chun, Talk at this Conference.
\bibitem{Chliap} P.Chliapnikov, Talk at this Conference.
\bibitem{Xie} Q.-B. Xie, Talk at this Conference.
\bibitem{Chan} A.H.Chan, Talk at this Conference.
\bibitem{Azimov} Ya.I.Azimov {\it et al},\Journal{\ZPC}{27}{65}{1985}.
\bibitem{Ochs} W.Ochs, Talk at this Conference.
\bibitem{Metzger} W.J.Metzger, Talk at this Conference.
\bibitem{Dokshit} Yu.L.Dokshitzer {\it et al}, \Journal{\NPB}{511}{396}{1998}.
\bibitem{Milstead} D.Milstead, Talk at this Conference.
\bibitem{Sterman} G.Sterman, Talk at this Conference.
\bibitem{Trel} D.Treleani, Talk at this Conference.
\bibitem{Calucci} G.Calucci, Talk at this Conference.
\bibitem{Pancheri} G.Pancheri, Talk at this Conference.
\bibitem{Ugoccioni} R.Ugoccioni, Talk at this Conference.
\bibitem{Walker} W.D.Walker, Talk at this Conference.
\bibitem{Feld} L.Feld, Talk at this Conference.
\bibitem{DeRoeck} A.DeRoeck, , Talk at this Conference.
\bibitem{Kim} V.Kim, Talk at this Conference.
\bibitem{Nikolaev} N.Nikolaev, Talk at this Conference.
\bibitem{Ivanov} D.Yu.Ivanov, R.Kirschner, \Journal{\PRD}{58}{114026}{1998}.
\bibitem{Pomeron} V.N.Gribov, {\it ZhETF} {\bf 41} {667} (1961);
 G.F.Chew and S.C.Frautschi, \Journal{\PRL} {7} {394} {1961};
\bibitem{review} A.Capella {\it et al}, {\it Phys. Rep.} {\bf 236} {225} (1994).\\
   A.Kaidalov, {\it Surveys in High Energy Physics} {\bf 13} {265} (1999).
\bibitem{1/N} G.t'Hooft, \Journal{\NPB} {72} {461} {1974};
 G.Veneziano, \Journal{\PLB} {52} {220} {1974}.
\bibitem{BFKL} L.N.Lipatov, {\it Sov.J.Nucl.Phys.} {\bf 23} {338} (1976);
E.A.Kuraev, L.N.Lipatov and V.S.Fadin, {\it  Sov.Phys.JETP} {\bf 45} {199} (1977);
Ya.Balitskii and L.N.Lipatov, {\it Sov.J.Nucl.Phys.} {\bf28} {822} (1978).
\bibitem{Anderson} B.Andersson, G.Gustafson and J.Samuelson, \Journal{\NPB}
{467} {443} {1996}.
\bibitem{Kancheli} L.P.A.Haakman, O.V.Kancheli, J.Koch, \Journal{\PLB} {391} {157} {1997}.
\bibitem{Fadin} V.S.Fadin and L.N.Lipatov, \Journal{\PLB} {429} {127} {1998}
\bibitem{Camici} M.Ciafaloni and G.Camici, \Journal{\PLB} {430} {349} {1998};
\bibitem{Brodsky} S.Brodsky {\it et al}, {\it JETP Lett.}{\bf 70}{155}(1999).
\bibitem{Ciafaloni} M.Ciafaloni {\it et al.}, hep-ph/9905566,9907409.
\bibitem{Colferai} D.Colferai, Talk at this Conference.
\bibitem{White} A.White, Talk at this Conference.
\bibitem{Balit} Ya.Balitsky, Talk at this Conference.
\bibitem{McLer} L.McLerran, R.V.Venugopalan, \Journal{\PLB} {424} {15} {1998};
 L.McLerran, hep-ph/9903536.
\bibitem{Jalil} J.Jalilian-Marian, Talk at this Conference.
\bibitem{Venug} R.Venugopalan, Talk at this Conference.
\bibitem{Fried} H.Fried, Talk at this Conference.
\bibitem{Simonov} A.Kaidalov and Yu.Simonov, in preparation.
\bibitem{Brower} R.Brower, talk at this Conference.
\bibitem{Gotsman} E.Gotsman, Talk at this Conference.
\bibitem{Ducati} B.Gay Ducati, Talk at this Conference.
\bibitem{Piotr} K.Piotrzkowski, talk at this Conference.
\bibitem{Haut} F.Hautman, Talk at this Conference.
\bibitem{Kaid} A.Kaidalov, Talk at this Conference.
\bibitem{Convery} M.Convery, Talk at this Conference.
\bibitem{Barreto} J.Barreto, Talk at this Conference.
\bibitem{Goul} K.Goulianos, Talk at this Conference.
\bibitem{Tan} Chung-I Tan, {\it Phys.Rep.} {\bf 315} {175} (1999).
\bibitem{Sarc} I.Sarcevich, Talk at this Conference.
\bibitem{Braun} M.Braun, Talk at this Conference.
\bibitem{Hwa} R.Hwa, Talk at this Conference.
\bibitem{Engel} R.Engel, Talk at this Conference.
\bibitem{Merino} C.Merino, Talk at this Conference.
\bibitem{Stasto} A.Stasto, Talk at this Conference.

\end{thebibliography}
\end{document}